\documentclass[%
reprint, 
superscriptaddress,
 amsmath,amssymb,
 aps, physrev,
]{revtex4-2}

\usepackage{graphicx}% Include figure files
\usepackage{dcolumn}% Align table columns on decimal point
\usepackage{bm}% bold math
\usepackage[utf8]{inputenc}
\usepackage[T1]{fontenc}
\usepackage{mathptmx}
\usepackage{etoolbox}
\usepackage{color}
\usepackage{ulem}
\usepackage{siunitx}
\usepackage[percent]{overpic}

\usepackage[acronym, shortcuts]{glossaries}
\newacronym{BSG}{BSG}{borosilicate glass}
\newacronym{CPT}{CPT}{coherent population trapping}
\newacronym{UV}{UV}{ultraviolet}
\newacronym{VCSEL}{VCSEL}{vertical-cavity surface-emitting laser}

\begin{document}

\preprint{APS/Phys. Rev. Appl.}

\title{\textbf{Microfabricated alkali vapor cells with tunable He-Ne buffer gas mixture using reservoirs with laser-actuated break-seals} 
}

\author{Cl\'ement Carl\'e}
\affiliation{Universit\'e Marie et Louis Pasteur, SUPMICROTECH, CNRS, institut FEMTO-ST, Besançon, France}
\author{Andrei Mursa}
\affiliation{Universit\'e Marie et Louis Pasteur, SUPMICROTECH, CNRS, institut FEMTO-ST, Besançon, France}
\author{Gabriel Faure}
\affiliation{Universit\'e Marie et Louis Pasteur, SUPMICROTECH, CNRS, institut FEMTO-ST, Besançon, France}
\affiliation{Tronics Microsystems, 98 Rue du Pré de l'Horme, 38926 Crolles, France}
\author{Shervin Keshavarzi}
\affiliation{Universit\'e Marie et Louis Pasteur, SUPMICROTECH, CNRS, institut FEMTO-ST, Besançon, France}
\author{Quentin Tanguy}
\affiliation{Universit\'e Marie et Louis Pasteur, SUPMICROTECH, CNRS, institut FEMTO-ST, Besançon, France}
\author{Emmanuel Klinger}
\affiliation{Universit\'e Marie et Louis Pasteur, SUPMICROTECH, CNRS, institut FEMTO-ST, Besançon, France}
\author{Vincent Maurice}
\affiliation{Univ. Lille, CNRS, Centrale Lille, Univ. Polytechnique Hauts-de-France, UMR 8520 - IEMN - Institut d’Electronique de
Microélectronique et de Nanotechnologie, Lille, France}
\author{Rodolphe Boudot}
\affiliation{Universit\'e Marie et Louis Pasteur, SUPMICROTECH, CNRS, institut FEMTO-ST, Besançon, France}
\author{Nicolas Passilly}
\email{Contact author: nicolas.passilly@femto-st.fr}
\affiliation{Universit\'e Marie et Louis Pasteur, SUPMICROTECH, CNRS, institut FEMTO-ST, Besançon, France}

\date{\today}% It is always \today, today,

\begin{abstract}
This letter reports on the generation of a tunable buffer gas mixture within microfabricated alkali vapor cells. We show that the combination of low-permeation windows with sequential openings of laser-actuated break-seals enables adjustment of a helium-neon (He-Ne) noble gas mixture, fully compatible with alkali metal dispensers. The gas reservoirs and the main cell cavities are initially sealed at the wafer level under distinct helium and neon atmospheres, respectively. Within each cell, after Cs vapor is released from the dispenser, the break-seals are successively actuated to incrementally increase the helium fraction in the buffer gas mixture. This process shifts the atomic clock frequency turnover temperature toward higher values. As an illustration, one of the fabricated cells was operated at 95$^{\circ}$C in a coherent population trapping clock, achieving a fractional frequency stability of 9$\times$10$^{-11}$ at one-day integration time. These results demonstrate the feasibility of precisely tuning buffer gas compositions in microfabricated vapor cells and support the suitability of He-Ne mixtures for miniature atomic clock applications.
\end{abstract}

\maketitle

%\tableofcontents

\section{\label{sec:level1}Introduction}

Microfabricated alkali vapor cells are fundamental to high-performance and wafer-scalable chip-scale atomic devices (CSADs)~\cite{Kitching:APR:2018}, such as clocks~\cite{Knappe:APL:2004, Batori:PRap:2023, Carle:OE:2023}, magnetometers~\cite{Shah:Nature:2007, Lucivero:OE:2022}, microwave-field sensors~\cite{Zhu:IEEEED:2022, Yuan:RPP:2023}, or voltage references~\cite{Teale:AVS:2022}. They also serve as outstanding platforms for conducting atomic spectroscopy experiments~\cite{Pate:OL:2023, Klinger:OL:2024}, quantum memories~\cite{Treutlein:2023} or laser-cooling demonstrations~\cite{McGilligan:APL:2020}. 

Microfabricated cells usually consist of one or two cavities etched in silicon, filled with alkali vapor, and sandwiched between two anodically bonded glass wafers. The silicon cavities can be structured by dry~\cite{Kitching:APL:2002} or wet~\cite{Chutani:ScRep:2015} etching, ultrasonic drilling~\cite{Woetzel:RSI:2011}, or water-jet cutting~\cite{Dyer:JAP:2022}. Additionally, various methods have been explored to fill the cell with alkali vapor, each with its own advantages and disadvantages. To prevent alkali consumption and interaction with glass during anodic bonding, former methods involving the insertion of pure metal have been set aside in favor of those that are based on the decomposition  of stable compounds into elemental alkali metal after sealing. They include the use of alkali azides~\cite{Liew:APL:2007, Dyer:JAP:2022}, chlorides~\cite{Knappe:OL:2005, Bopp:JPP:2021}, chromates  or molybdates dispensers in the form of solid pills~\cite{Douahi:EL:2007, Vicarini:SA:2018} or paste~\cite{Maurice:APL:2017}, as well as graphite reservoirs~\cite{Kang:APL:2017, Martinez:NC:2023}.

In CSADs, the cell is usually filled with a buffer gas. Its presence reduces the wall-collision relaxation rate, which enhances the observed atomic coherence lifetime, and thereby improves the stability or sensitivity of the device. In addition, a precise control of the buffer gas pressure in the cell is crucial.
In magnetometers, the relaxation time of spin-polarized atoms is optimized at a specific buffer gas pressure~\cite{McWilliam:PRAppl:2024, Shah:Nature:2007} and fluorescence quenching can benefit from gas mixtures. In microwave cell clocks, the buffer gas pressure induces a collisional shift of the hyperfine transition frequency that can jeopardize long-term stability. A mixture of two buffer gases, each of them leading to opposite shifts of the clock frequency, is therefore frequently used to cancel the temperature dependence at a specific target turnover temperature~\cite{Vanier:JAP:1982, Kozlova:PRA:2011}, whose value is fixed by the ratio $a = P_1/P_2$, where $P_i$ is the buffer gas pressure of the gas $i$ (with $i =$ 1,2).

Filling a microfabricated vapor cell with buffer gas is typically achieved by backfilling the bonder chamber with the desired gases before sealing the cell. However, this method has certain drawbacks, including potential inconsistencies in gas pressure between different cells on a wafer and the limitation that all cells produced from the same wafer must have the same target buffer gas pressure. For instance, a strategy has recently been introduced in which rubidium is generated in situ from the thermal decomposition of a BaN$_{6}-$RbCl precursor, while a controlled Ar/N$_2$ buffer gas mixture is simultaneously sealed into the wafer~\cite{Li2025}. This approach highlights the possibility of integrating alkali generation and buffer gas introduction within the same fabrication process, although the precursor decomposition makes precise control of the buffer gas ratio challenging. Another approach employs alkali azides whose \ac{UV} decomposition releases both the alkali and nitrogen. In principle, one could control the amount of nitrogen by adjusting the \ac{UV}-irradiation. However, the amount of nitrogen is directly tied to the quantity of alkali metal released, as determined by the azide stoichiometry, which restricts the range of achievable nitrogen pressures. Additionally, this method suffers from significant variability in decomposition yield~\cite{Dyer:JAP:2022}. Similarly, controlling the depletion of nitrogen buffer gas through the activation of getter compounds in an alkali pill dispenser~\cite{Dyer:APL:2023} ties the quantities of alkali metal and nitrogen together. Furthermore, both decomposition and adsorption processes are expected to continue at a small rate even without illumination. While this may be compatible with sensors such as optically-pumped magnetometers, it could limit the long-term stability of atomic clocks.

It therefore appears more reliable to combine the high yield of alkali dispensers with a mixture of noble gases, as these gases do not react with the getter compound. 
In this context, several noble gas mixtures can be considered, such as Ne–Ar and Ne–He. In the former case, the addition of argon to neon lowers the clock transition turnover temperature~\cite{Boudot2011}, whereas in the latter, the addition of helium raises it~\cite{Kroemer:OE:2015}. Such complementary effects provide a convenient way to tailor the operating temperature of microfabricated vapor cells to the requirements of miniature atomic clock applications.
However, since mono-atomic species are likely to leak, the cell hermeticity must be addressed. To achieve this goal, we have recently investigated the use of aluminosilicate glass, instead of borosilicate, along with alumina coatings, to mitigate gas permeation~\cite{Carle:JAP:2023, Carle:JAP:2024}. Based on these findings, helium, initially employed as a tracer gas, has emerged as a viable candidate for a buffer gas, at least in a mixture.

In addition, we proposed an approach to fill and seal microfabricated vapor cells, inspired from glass-blowing techniques~\cite{Maurice:NMN:2022}. In this initial demonstration, cells with evacuated main cavities and four gas reservoirs filled with neon were used. Sequential laser ablation of the silicon walls resulted in an incremental increase in neon pressure. Notably, measurements taken after only two reservoirs were opened indicated minimal neon leakage from the intact reservoirs into the science cavity (where the atom-light interaction occurs), i.e. on the order of 10$^{-4}$ Torr/day. This implies that it is not mandatory to open all the reservoirs; instead, they can be selectively opened to adjust the pressure as needed.

In this work, we report on the design, development and characterization of Cs vapor microfabricated cells that can be filled with a tunable He-Ne buffer gas mixture. This tunability is achieved through the use of multiple gas reservoirs. Whereas the science and dispenser cavities are initially prefilled with a fixed pressure of neon, controlled amounts of helium gas are gradually introduced by sequentially opening several helium reservoirs located around the science cavity (Fig.~\ref{fig:1}). The amount of added helium, and consequently the mixture ratio, is then adjusted in a manner analogous to a potentiometer, by varying the number of opened reservoirs. This is experimentally validated by the progressive shift of the clock frequency turnover point, which in this work has been successfully set at temperatures as high as 100$^{\circ}$C. A \ac{CPT} clock, utilizing such a cell with an approximately 4.5\% helium-based mixture, achieves a fractional frequency stability of 9~$\times$~10$^{-11}$ at 1~day.

\section{\label{sec:level2}Cell microfabrication}

\begin{figure}[t!]
\centering
\includegraphics[width=0.9\linewidth]{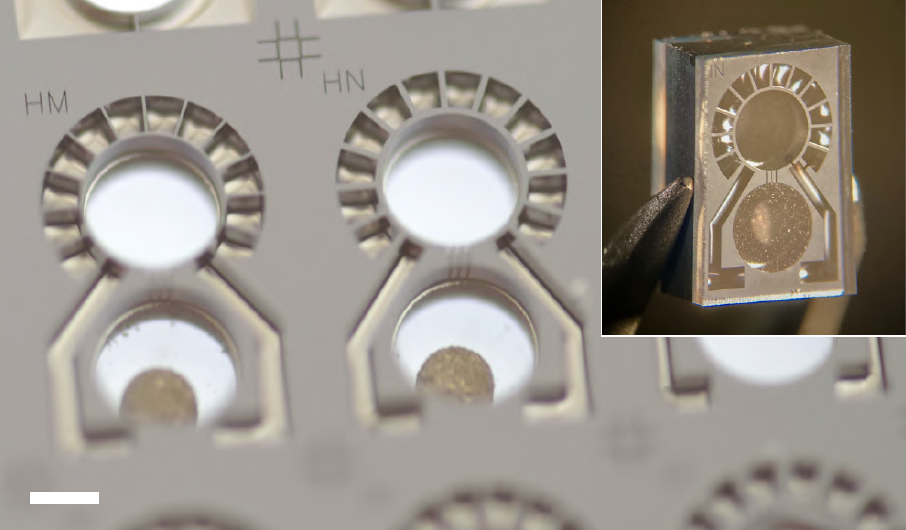}
\caption{Photograph of a wafer of microfabricated cells. The multiple small-size gas reservoirs surrounding the science cavity are visible. The lower cavity holds the pill dispenser. The white bar is 1~mm long. Inset: A single cell after dicing and activation of two break-seals to connect the reservoirs furthest to the right.}
\label{fig:1}
\end{figure}

The buffer-gas filled Cs vapor cells developed in this study rely on a similar cell technology as described in Refs~\cite{Douahi:EL:2007, Vicarini:SA:2018, Maurice:NMN:2022}, based on a stack of silicon and glass wafers, and alkali dispensers. Here, the different cavities are structured within the silicon wafer by double-sided lithography and two-step deep reactive ion etching. This process etches first all cavities to nearly half of the depth of the Si wafer. Hence, a second etching step is performed onto the other Si side to etch-through only the main cavities, but not the reservoirs. This approach allows trapping helium during the first anodic bonding performed between the reservoirs side of the Si wafer and an aluminosilicate glass wafer. After loading the pill dispensers (trade name Cs/AMAX/Pill/1-0.6 from SAES Getters), a second anodic bonding is performed under neon atmosphere to fill the main cavities. The resulting cells are shown in Fig.~\ref{fig:1}. They feature two main cavities: the lower cavity, which houses the pill dispenser, and the upper one, where the atom-light interaction occurs. These cavities are connected by narrow channels and initially filled with only neon. Each cavity has a diameter of \SI{2}{\milli\meter} and a length of either \SI{1}{\milli\meter} or \SI{1.5}{\milli\meter}, depending on the fabricated wafer. Therefore, up to 14 helium-filled reservoirs are positioned around the science cavity. These reservoirs are either \SI{600}{\micro\meter} or \SI{750}{\micro\meter} deep, depending on the fabricated wafer, which constitute approximately 1.5\% of the total volume of the cell. In some of the cells (as in Fig.~\ref{fig:1}), two larger reservoirs, each corresponding approximately to 8\% of the cell volume, have been etched in order to establish an initial gas mixture ratio. This allows for finer adjustments with the smaller reservoirs, which are all identical in volume. Once the dispenser is activated by laser local-heating to release the Cs vapor, the \SI{100}-\SI{150}{\micro\meter}-thick Si walls separating the reservoirs from the science cavity, can be processed by laser ablation. 

\section{\label{sec:level3}Characterization setup}

Microfabricated cells have been characterized with a \ac{CPT} atomic clock setup, comparable to the one described in Ref.~\cite{Carle:OE:2023}. Here, the laser source is a \ac{VCSEL}, tuned on the Cs D$_1$ line at 895~nm. Its current is modulated at \SI{4.596}{\giga\hertz} to generate two first-order optical sidebands frequency-split by \SI{9.192}{\giga\hertz} required for \ac{CPT} interaction. The microwave signal is provided by a microwave synthesizer referenced by a hydrogen maser, used as a reference for frequency shifts and frequency stability measurements. At the output of the \ac{VCSEL}, an acousto-optic modulator is used as an optical shutter to produce the symmetric Auto-Balanced Ramsey (SABR) pulsed interrogation sequence~\cite{MAH:APL:2018}. This technique was demonstrated to mitigate light-shifts (sensitivity of the clock frequency to laser power, laser frequency and microwave power) by more than two orders of magnitude~\cite{MAH:APL:2022}. The laser beam finally passes through the microfabricated cell before being detected by a photodiode. The extracted atomic signal is then processed by electronics for laser frequency stabilization, local oscillator frequency stabilization onto the center of the central Ramsey-\ac{CPT} fringe and light-shift compensation~\cite{MAH:APL:2022}. The cell is temperature controlled into a physics package covered by a mu-metal magnetic shield. A static magnetic field of 234~mG is applied to isolate the clock transition.

In these conditions, the measured clock frequency $\nu_0$ is given by
\begin{equation}
\nu_0 = \nu_{Cs} + \Delta \nu_{bg} + \Delta \nu_l + \Delta \nu_{z},
\label{eq:shift}
\end{equation}
with $\nu_{Cs}$~=~\SI{9192631770}{\hertz} being the Cs atom unperturbed clock frequency, $\Delta \nu_{bg}$ the collisional shift, $\Delta \nu_l$ the light shift, and $\Delta \nu_{z}$ the Zeeman shift. In our setup, we have $\Delta \nu_{z}~\simeq$~\SI{23.4}{\hertz} and we neglect the light-shift $\Delta \nu_l$ ($\Delta \nu_l$~=~0) due to the use of the SABR sequence. Equation~(\ref{eq:shift}) then simplifies to $\nu_0~=~\nu_{Cs}~+~\Delta \nu_{bg}$. In a buffer gas-filled cell, the collisional shift is well-approximated in a limited temperature range by
\begin{equation}
\Delta \nu_{bg} = P [(\sum r_i\beta_i) + (\sum r_i\delta_i) (T - T_0) 
+ (\sum r_i\gamma_i) (T - T_0)^2],
\label{eq:shiftcoll}
\end{equation}
with $P$ being the total buffer gas pressure, $T$ the cell temperature, $T_0$~=~273.16 K the reference temperature, $\beta_i$, $\delta_i$ and $\gamma_i$, with $i$~=~1,2, being the pressure and temperature coefficients of buffer gas $i$. Additionally, $r_i$ represents the ratio between the buffer gas $i$ pressure and the total pressure, such that $r_1 + r_2$~=~1. Buffer gas coefficients reported in Ref.~\cite{Kozlova:PRA:2011} for Ne and in Ref.~\cite{Beverini:OC:1981} for He were utilized in this study. In a buffer gas mixture, the temperature dependence of the frequency shift vanishes at the inversion temperature $T_i$ for the pressure ratio $a = P_2/P_1$ given by
\begin{equation}
a = - \frac{\delta_1 + 2 \gamma_1 (T_i - T_0)}{\delta_2 + 2 \gamma_2 (T_i - T_0)}.
\label{eq:pressratio}
\end{equation}
Therefore, the methodology for cell characterization involves the following steps. Initially, a cell filled only with neon (with no reservoirs opened) is placed in the clock setup. An automated routine is then used to gradually increase the cell temperature from 60$^{\circ}$C to 90-100$^{\circ}$C, in 2$^{\circ}$C increments [Fig.~\ref{fig:2}(a)]. At each temperature step, the clock frequency is acquired for 10 minutes and the average value is extracted for each cell temperature set point. After completing this initial run with pure neon, helium reservoirs are progressively opened to release helium gas into the science cavity. The same characterization procedure is then repeated after each series of reservoir openings.

\section{\label{sec:level4}Measurements of frequency temperature dependence}

\begin{figure}[ht!]
\centering
\includegraphics[width=0.95\linewidth]{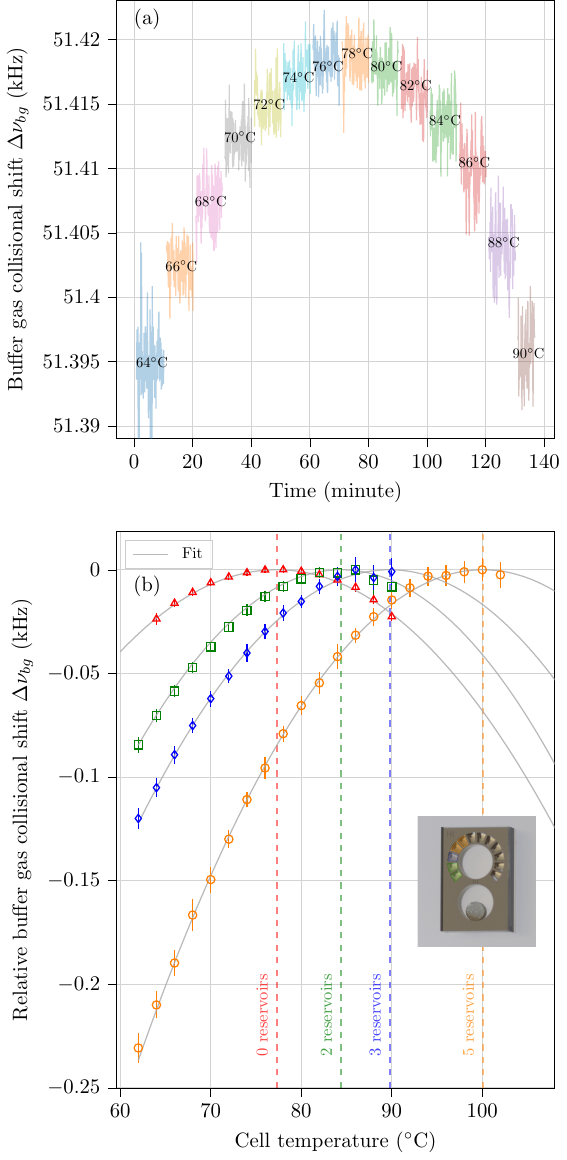}
\caption{(a) Example of temporal trace of the clock frequency for different cell temperatures in cell A, prior to any opening of the helium gas reservoirs, i.e., when the science cavity is filled only with neon. Mean values and standard deviations are extracted from each temperature step. (b) Buffer gas collisional shift $\Delta \nu_{bg}$ versus the cell temperature before (pure neon configuration, red triangles) and after the opening of 2 (green squares),  3 (blue diamonds) and 5 (orange circles) helium gas reservoirs. Opened reservoirs are highlighted by colored areas on the microcell photograph shown in the inset.}
\label{fig:2}
\end{figure}

Figure~\ref{fig:2}(a) displays the typical temporal trace of the clock frequency for a tested cell recorded before any helium reservoirs were opened (using pure neon), across temperatures ranging from 64$^{\circ}$C to 90$^{\circ}$C. Derived from these data, Fig.~\ref{fig:2}(b) reports the corresponding temperature dependence of the collisional shift $\Delta \nu_{bg}$, in pure neon configuration (red data). Experimental data have been non-linearly fitted with the orthogonal distance regression algorithm~\cite{Boggs:1989} using a polynomial function to extract the inversion temperature $T_i$~=~77.3~$\pm$~0.3$^{\circ}$C. This value is in good agreement with previously reported turnover temperatures $T_i$ for a Cs cell filled with neon~\cite{Kozlova:PRA:2011, Kozlova:TIM:2011}. 

To introduce helium into the science cavity, an initial set of two helium reservoirs, highlighted in green in the inset of Fig.~\ref{fig:2}(b), was opened. The presence of helium was confirmed by two observations, firstly, an increase in the absolute value of the clock frequency, indicating a rise in He buffer gas pressure (although the figure displays relative frequency values for clarity), and secondly, an increase in the turnover temperature~\cite{Kroemer:OE:2015} measured at $T_i$ = 84.4$~\pm~$0.7$^{\circ}$C. Subsequently, a third reservoir, marked in blue in Fig.~\ref{fig:2}, was opened, further shifting the turnover temperature to 89.8~$\pm$~1.8$^{\circ}$C. Finally, two additional reservoirs were activated, marked in orange in Fig.~\ref{fig:2}, raising the turnover temperature to 100.1~$\pm$~1.0$^{\circ}$C. It is worth noting that after each reservoir is opened, a delay is required for the newly introduced gas to fully diffuse into the main cavity. Unlike a vacuum cavity filled with a single gas, as in Ref.~\cite{Maurice:NMN:2022}, the mixing dynamics of two gases are slower, and equilibrium in our case took up to several weeks following the final reservoir openings. This prolonged mixing time is likely due to the small diameter of the ablated channels, which are moreover arranged sequentially, as they are positioned between the reservoirs to prevent contamination of the main cavity with ablation debris. Furthermore, each reservoir opening introduces an additional internal volume that must be filled with cesium vapor. Toward the end of the sequence, this likely reduced the atomic density in the main cavity. Once we observed that the frequency shift began accelerating —even though the gas composition was presumably stabilized— we reactivated the dispenser. Working with inert buffer gases that do not chemically interact with the dispenser makes such corrective reactivation possible when needed.

\begin{figure}[t!]
\centering
\includegraphics[width=0.95\linewidth]{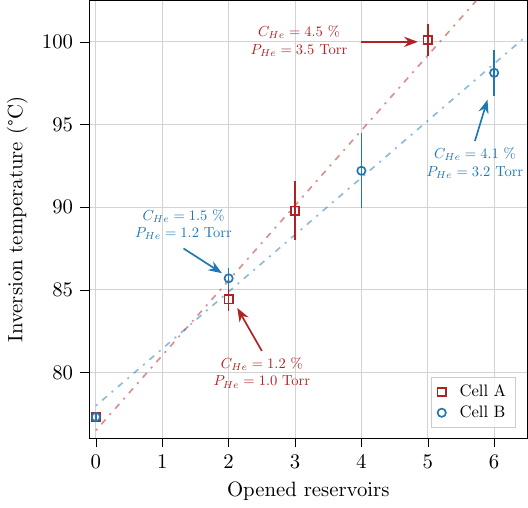}
\caption{Inversion temperature as a function of the number of opened reservoirs for two microcells, A and B. The difference in slopes between the two cells is attributed to a slight difference in reservoir volume: Reservoir walls are \SI{100}{\micro\meter} thick in cell A and \SI{125}{\micro\meter} thick in cell B. The laser-actuated valve opening acts as a potentiometer, enabling fine tuning of the buffer gas mixture pressure ratio. Buffer gas coefficients from Ref.~\cite{Kozlova:PRA:2011} (Ne) and Ref.~\cite{Beverini:OC:1981} (He) were considered.}
\label{fig:3}
\end{figure}

Figure~\ref{fig:3} summarizes how the sequential opening of helium gas reservoirs enables precise tuning of Ne-He gas mixtures in Cs cells. Let us first consider the cell A, previously shown in Fig.~\ref{fig:2}. The two first opened reservoir units (marked in green in Fig.~\ref{fig:2}(b)) correspond to a relative volume of $r_v = V_r / (V_r + V_{sc}) = 4.2~\%$, where $V_r$ is the reservoir volume and $V_{sc}$ is the science cavity volume (including the dispenser cavity). This configuration yields an inversion temperature of $T_i$ = 84.4~$\pm$~0.7$^{\circ}$C, which, using the coefficients from Refs.~\cite{Kozlova:PRA:2011, Beverini:OC:1981}, corresponds to an estimated Ne-He mixture containing 1.2~\% of helium. Using the same method, the opening of a third reservoir (marked in blue in Fig.~\ref{fig:2}(b)) increased the relative volume to $r_v$ =~6.2~\%, leading to an estimated He concentration of 2.3~\%. Finally, the opening of two additional reservoirs, (highlighted in orange in Fig.~\ref{fig:2}(b)), results in an inversion temperature of $T_i$ = 100.1~$\pm$~1.0$^{\circ}$C, close to the target value and corresponding to a helium concentration of 4.5~\%.

The experimental data points for the inversion temperature exhibit a nearly linear dependence on the relative volume of opened reservoirs and align well with the theoretical model considering an initial helium pressure of nearly 36~Torr at 70$^{\circ}$C. The corresponding curves, shown as the dashed lines in Fig.~\ref{fig:3}, can thus be used to predict the number of reservoirs that need to be opened to achieve a specific target inversion temperature. The second cell, denoted B, shows a similar trend upon the sequential opening of 6 reservoirs, albeit with a lower slope than that of cell A. This is mostly attributed to the reduced volume of the reservoirs,  due to thicker separation walls (\SI{125}{\micro\meter} in B vs. \SI{100}{\micro\meter} in A). Nevertheless, opening 6 reservoirs brings the inversion temperature close to 100$^{\circ}$C. It is also worth noting that when the cell includes a larger initial reservoir (as shown in Fig.~\ref{fig:1}), its opening first induces a significant shift in the inversion temperature — as observed in one cell, where it increased from 78$^{\circ}$C to 96$^{\circ}$C. Consequently, by lowering the helium pressure during fabrication, finer control over the final helium concentration can be achieved through the successive opening of the smaller surrounding reservoirs.

\section{\label{sec:level5}Frequency stability measurements}

\begin{figure}[t!]
\centering
\includegraphics[width=0.95\linewidth]{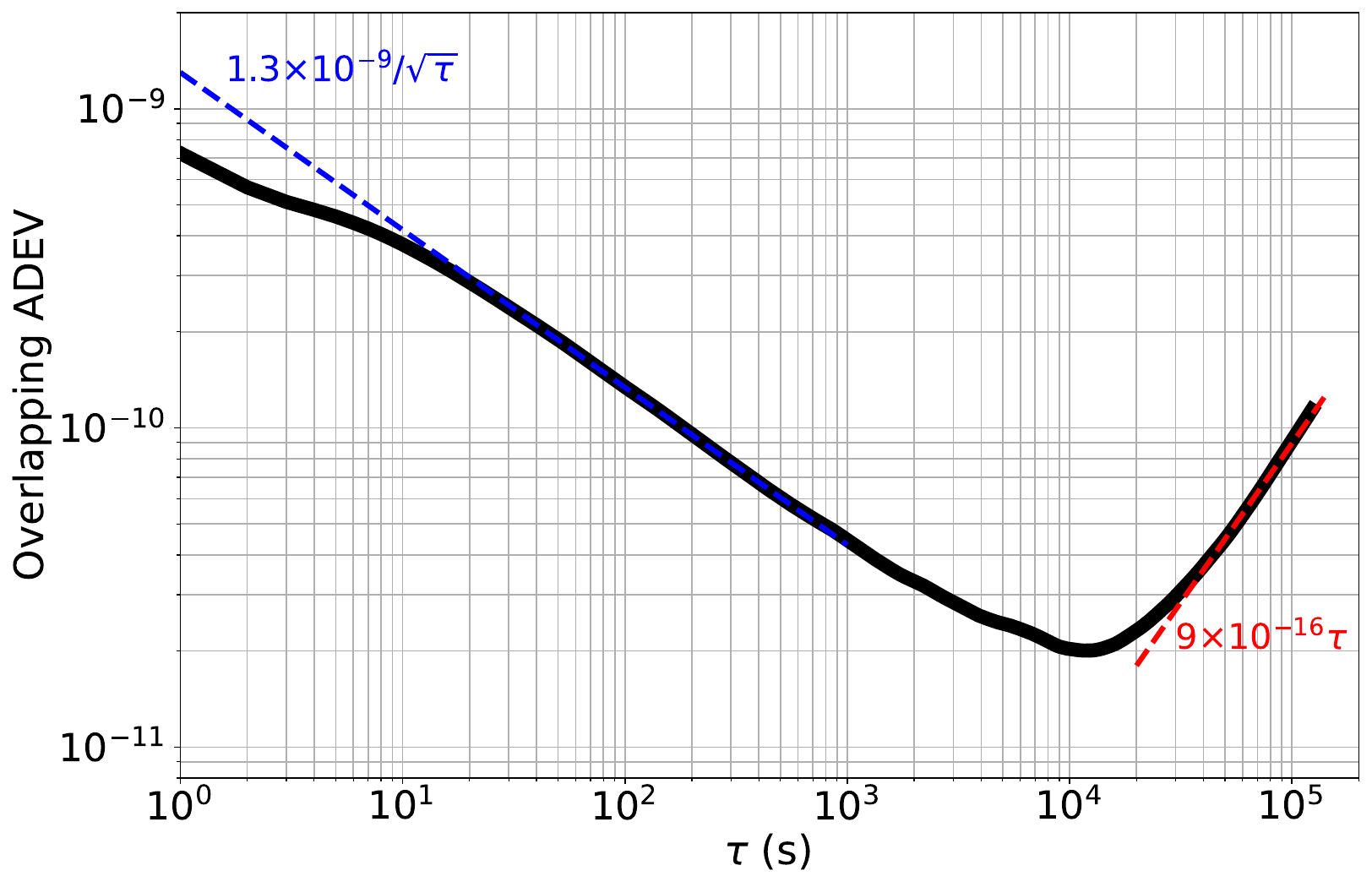}
\caption{Allan deviation of a CPT clock  based on a Cs-He-Ne microfabricated cell maintained around 95°C.}
\label{fig:4}
\end{figure}

To validate the suitability of cells combining low-permeation windows with a He-Ne buffer gas mixture for clock operation, we eventually conducted clock frequency stability measurements using one of the Cs-He-Ne microcell (cell A). This cell, featuring a 1~mm-long cavity, was temperature stabilized at only 95$^{\circ}$C, to minimize optical absorption while maintaining sufficient signal contrast. The resulting clock Allan deviation, shown in Fig.~\ref{fig:4}, exhibits a fractional frequency stability of 7.0~$\times$~10$^{-10}$ at 1~s and 9.0~$\times$~10$^{-11}$ at 10$^5$~s. These long-term stability values are consistent with expectations based on helium permeation through uncoated aluminosilicate glass. Nevertheless, further improvements could potentially be achieved using additional alumina coatings~\cite{Carle:JAP:2023}. For temperatures above 90~$^{\circ}$C, the fractional thermal sensitivity ($1/\nu_0 \times d\Delta \nu_{bg}/dT$) is significantly reduced in a cell containing a buffer gas mixture with 4.5\% helium, compared to a cell filled solely with $\approx$93~Torr of neon. Specifically, at 95$^{\circ}$C, the reduction reaches a factor of three (from -5.4$\times$10$^{-10}$~K$^{-1}$ down to 1.8$\times$10$^{-10}$~K$^{-1}$ based on the coefficients from~\cite{Kozlova:PRA:2011}) and as much as a factor of 40 at 100$^{\circ}$C. In practical terms, to keep the frequency instability contribution below 1.0$\times$10$^{-11}$, a pure neon cell requires temperature stabilization within $\pm$19~mK at 95$^{\circ}$C and $\pm$14~mK at 100$^{\circ}$C. In contrast, the He-Ne mixture relaxes this constraint to $\pm$56~mK at 95$^{\circ}$C and up to $\pm$565~mK at 100$^{\circ}$C.
\\

\section{\label{sec:level6}Conclusion}
In conclusion, we have demonstrated the effectiveness of a filling technique that leverages laser-actuated break-seal gas reservoirs to microfabricate cells with a tunable He-Ne buffer gas mixture. The helium-to-neon pressure ratio in the cell's science cavity can be precisely adjusted by tailoring the relative reservoir and cavity volumes. The successful introduction of helium into cells initially pre-filled with neon was confirmed by the observed increase in the clock frequency turnover point, reaching temperatures up to approximately 100$^{\circ}$C. In addition, one of the fabricated cell was operated in a coherent population trapping clock and exhibited a long-term fractional frequency stability of 9.0~$\times$~10$^{-11}$ at one day, supporting the suitability of the He-Ne buffer gas mixture when cell hermeticity is enhanced by low-permeation windows. This approach holds promise for the development of microfabricated vapor cells with precisely controlled buffer gas compositions, adaptable to various alkali species. 

\begin{acknowledgments}
This work was supported by the Direction G\'{e}n\'{e}rale de l’Armement (DGA) and by the Agence Nationale de la Recherche (ANR) in the frame of the ASTRID project named PULSACION (Grant ANR-19-ASTR-0013-01), LabeX FIRST-TF (Grant ANR 10-LABX-48-01), EquipX Oscillator-IMP (Grant ANR 11-EQPX-0033) and EIPHI Graduate school (Grant ANR-17-EURE-0002). The PhD thesis of C. Carl\'e was co-funded by Centre National d'Etudes Spatiales (CNES) and Agence Innovation D\'efense (AID). This work was supported by the french RENATECH network and its FEMTO-ST technological facility (MIMENTO). The authors want to thank Femto-Engineering, in particular J. Safioui and E. Elmir, for the support with laser ablation as well as A. Bresson from MIMENTO.
\end{acknowledgments}

\section*{Data availability statement}
The data supporting the findings of this study are available from the corresponding author upon reasonable request.
%\section*{Conflict of interest}
%The authors state that there is no conflict of interest to disclose.\\

\bibliography{CARLE_APS}% Produces the bibliography via BibTeX.

\end{document}